\documentclass{aa}
\usepackage{graphicx}

\begin{document}

\title{On the maximum rotational frequency of neutron and hybrid stars}

\author{G. F. Burgio\inst{1} \and H.-J. Schulze\inst{1} \and F. Weber\inst{2}}

\institute{Istituto Nazionale di Fisica Nucleare, Sezione di Catania, Via S.
  Sofia 64, I-95123 Catania, Italy \and Department of Physics, San Diego State
  University, 5500 Campanile Drive, San Diego, CA 92182, USA}

\date{Received / Accepted}


\abstract{We construct self-consistent equilibrium sequences of general
  relativistic, rotating neutron star models. Special emphasis in put on the
  determination of the maximum rotation frequency of such objects.  Recently
  proposed models for the equation of state of neutron star matter are
  employed, which are derived by describing the hadronic phase within the
  many-body Brueckner--Bethe--Goldstone formalism, and the quark matter phase
  within the MIT bag model using a density dependent bag constant.  We find
  that the rotational frequencies of neutron stars with deconfined quark
  phases in their cores rival those of absolutely stable, self-bound
  strange quark matter stars.  This finding is of central importance for the
  interpretation of extremely rapidly rotating pulsars, which are the targets
  of present pulsar surveys.  }

\titlerunning{On the maximum rotational frequency....}

\maketitle

\section{Introduction}

Bulk properties of neutron stars depend sensitively on the equation of state
(EOS) of nuclear matter at supernuclear density (Shapiro \& Teukolsky 1983).
In fact, matter in the cores of neutron stars possesses densities ranging from
a few times $n_0$ $(\approx 0.17\;{\rm fm}^{-3}$, the normal nuclear matter
density) up to one order of magnitude higher. The EOS at such extreme
densities is only poorly known. In the simplest model, a neutron star consists
of neutrons only. In a slightly more accurate representation, a neutron star
will contain neutrons and a small number of protons whose charge is balanced
by leptons.  Because of the extreme densities in the interior of a neutron
star, the neutron chemical potential will exceed the mass (modified by
interactions) of various members of the baryon octet. So in addition to
neutrons, protons, and electrons, neutron stars may be expected to have
populations of hyperons, too (Glendenning 1982, 1985). A comprehensive
description of neutron star matter should thus account not only for nucleons
and leptons, but also for hyperons, mesons, and the plausible transition to a
quark-gluon plasma state (Glendenning 1982; Baym et al.\ 1985;
  Glendenning 1985, 1990; Weber 1999a, 1999b; Glendenning 2000).

Because of the unprecedented advances in observational astronomy, one may hope
that measurements of neutron star parameters will soon constrain theories of
dense matter. The rotational periods of rapidly rotating neutron stars
(pulsars), for instance, provide restrictions on the EOS when combined with
the mass constraint. Radio pulsars are stable rotators which can be treated as
uniformly rotating bodies (Weber 1999a).  Their rotational frequencies are
bound from above by the Kepler frequency $\Omega_K$, which sets an absolute
limit on stable rotation because of mass shedding from the star's equator.
$\Omega_K$ depends sensitively on the EOS (Friedman et al.\ 1986; Friedman et
al.\ 1989; Lattimer et al.\ 1990; Cook et al.\ 1994; Salgado et al.\ 1994;
Weber 1999a, 1999b). An EOS which predicts Kepler frequencies that are smaller
than the observed rotational frequencies is to be rejected as it is not
compatible with observation.

In this paper we present results for configurations that are rotating
uniformly at their Kepler frequencies. Models for the EOS of dense neutron
star matter are computed for three different descriptions which are based on
the microscopic many-body Brueckner-Bethe-Goldstone (BBG) theory (Baldo 1999).
The first EOS describes hadronic matter with nucleons and leptons in
beta-equilibrium (Baldo et al.\ 1997). The second EOS includes the presence of
all hyperons ($\Sigma^-$ and $\Lambda$) that become populated in neutron star
matter up to the highest densities relevant for neutron stars (Baldo et al.\ 
2000a).  Finally, the third EOS accounts for a hypothetical phase transition
of confined hadronic matter into deconfined quark matter (Burgio et al.\ 
2002a, 2002b). The hadronic phase is described in the framework of the BBG
theory, and the deconfined quark phase within an extended 
MIT bag model (Chodos et
al.\ 1974). We use recent experimental results on the possible formation of a
quark-gluon plasma state (Heinz 2000) at CERN to constrain the bag constant,
$B$, which turns out to be density dependent.  

Calculations of neutron star (NS)
properties performed for such an EOS show that the maximum NS masses lie in
the relatively narrow interval of $1.4\, \rm M_\odot \leq M_{max} \leq 1.7\,
\rm M_\odot$. Moreover, we find that neutron stars
with deconfined quark matter in their interiors computed for an effective
density-dependent bag constant, can rotate at Kepler periods down to half a
millisecond. Such small rotational periods are not accessible to neutron stars
made of confined hadronic matter only nor to neutron stars with a hadron-quark 
phase
transition computed for a constant bag constant.  It was commonly believed
that only absolutely stable strange stars (Witten 1984), which are self-bound
objects, may rotate that rapidly (Glendenning 1990). Our findings,
however, suggest that pulsars rotating with about half a millisecond 
period could still
be interpreted as conventional neutron stars containing a metastable deconfined
quark phase at their centers.

This paper is organized as follows. In Section 2 we briefly illustrate the BBG
many-body theory. We then proceed to the discussion of the hadron-quark phase
transition described within the MIT bag model supplemented with a density
dependent bag constant.  The stellar structure equations are introduced in
Section 3. In Section 4 we present our results for both non-rotating as well as
for rotating neutron stars and discuss the impact of an effective, density
dependent bag constant on the Kepler frequency. Finally, conclusions are
drawn in Section 5.

\par
\section{Equations of state of neutron star matter}
We begin with the description of the hadronic phase, for which we have adopted
the Brueckner--Bethe--Goldstone (BBG) theory. This microscopic many-body
approach is based on a linked cluster expansion of the energy per nucleon of
nuclear matter (Baldo 1999).  The basic ingredient is the Brueckner reaction
matrix $G$, which is the solution of the Bethe--Goldstone equation,

\begin{equation}
G[n;\omega] = V  + \sum_{k_a k_b} V {{|k_a k_b\rangle  Q  \langle k_a k_b|}
  \over {\omega - e(k_a) - e(k_b) }} G[n;\omega], 
\label{e:BG}
\end{equation}                                                           
\noindent
where $V$ is the bare nucleon-nucleon (NN) interaction, $n$ is the nucleon
number density, and $\omega$ the starting energy.  The single-particle energy
$e(k)$ (assuming $\hbar$=1),
\begin{equation}
e(k) = e(k;n) = {{k^2}\over {2m}} + U(k;n),
\label{e:en}
\end{equation}
\noindent
and the Pauli operator $Q$ constrain the propagation of intermediate baryon
pairs above the Fermi momentum.  The Brueckner--Hartree--Fock (BHF)
approximation for the single-particle potential $U(k;n)$ using the {\it
  continuous choice} is
\begin{equation}
U(k;n) = {\rm Re} \sum _{k'\leq k_F} 
\langle k k'|G[n; e(k)+e(k')]|k k'\rangle_a,
\label{e:U}
\end{equation}
\noindent
where the subscript ``{\it a}'' indicates anti-symmetrization of the matrix
elements. Because of the occurrence of $U(k;n)$ in Eq.~(\ref{e:en}),
Eqs.~(\ref{e:BG}) through (\ref{e:U}) constitute a coupled system of equations
that need to be solved self-consistently for different densities (i.e., Fermi
momenta).  The basic input quantity in the Bethe-Goldstone equation
(\ref{e:BG}) is the NN interaction in free space, $V$. We adopt the elaborate
Paris potential (Lacombe et al.\ 1980) as a model for the free two-nucleon
interaction supplemented with the Urbana model UVII as three-body force
(Carlson et al.\ 1983; Schiavilla et al.\ 1986).  In the BHF approximation the
energy per nucleon is given by
\begin{equation}
{E \over{A}}  =  
          {{3}\over{5}}{{k_F^2}\over {2m}}  + {{1}\over{2n}} {\rm Re} 
\sum_{k,k'\leq k_F} \langle k k'|G[n; e(k)+e(k')]|k k'\rangle_a. 
\label{e:EA}
\end{equation}
\noindent
\begin{figure}
\centering
\resizebox{\hsize}{!}{\includegraphics[angle=270]{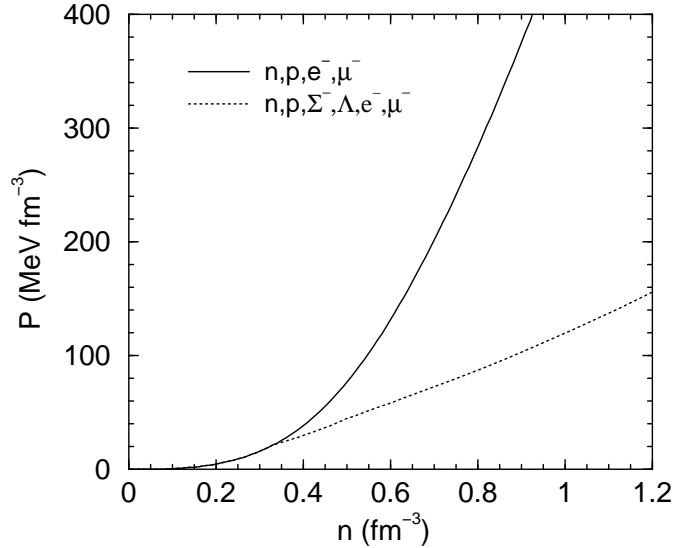}} 
\caption{BBG equation of state (EOS) of asymmetric, electrically charge 
  neutral hadronic matter in beta equilibrium. The solid line shows the EOS
  when only nucleons and leptons are present, whereas the dotted line shows
  the EOS when hyperons are included.}
\end{figure}

Higher order correlations have been demonstrated to be negligible over the
considered density range (Baldo et al.\ 2000b, 2001). The nuclear matter EOS
computed from Eqs.~(\ref{e:BG}) -- (\ref{e:EA}) fulfills several essential
requirements, namely (i) it reproduces the correct nuclear matter saturation
density, i.e., $n_0=0.176 \; {\rm fm}^{-3}$, (ii) the nuclear
incompressibility $K=281 \; {\rm MeV}$ is compatible with the values extracted
from phenomenology, (iii) the symmetry energy $a_{\rm sym} = 33.6 \; {\rm
  MeV}$ is compatible with nuclear phenomenology, and (iv) the causality
condition is strictly fulfilled (Baldo et al.\ 1997).
\par
Recently, we incorporated hyperons into the BBG treatment of neutron star
matter (Baldo et al.\ 1998, 2000a). This requires knowledge of the
nucleon-hyperon (NH) and hyperon-hyperon (HH) interactions. Because of the
lack of experimental data, the hyperon-hyperon interaction has been neglected
in a first approximation, whereas for the NH interaction the Nijmegen
soft-core model (Maessen et al.\ 1989) has been adopted. 

The equation of state of neutron star matter calculated along this scheme is
shown in Fig.\ 1.  The solid line displays the pressure as a function of baryon
number density of beta-stable, electrically charge neutral nuclear matter when
only nucleons and leptons are present.  The dashed line represents the
equation of state when $\Sigma^-$ and $\Lambda$ hyperons are self-consistently
included, too. It is evident that the presence of hyperons strongly softens the
equation of state (Glendenning 1982, 1985). This is caused by both the
larger number of baryonic degrees of freedom and the attraction felt by the
$\Sigma^-$ in the nuclear medium. As it turns out, this model for the EOS
produces a maximum neutron star mass that lies slightly below the 
canonical value of $1.44\, M_\odot$ (Taylor \& Weisberg 1989). 
\begin{figure*}
\centering
\resizebox{\hsize}{!}{\includegraphics[angle=270]{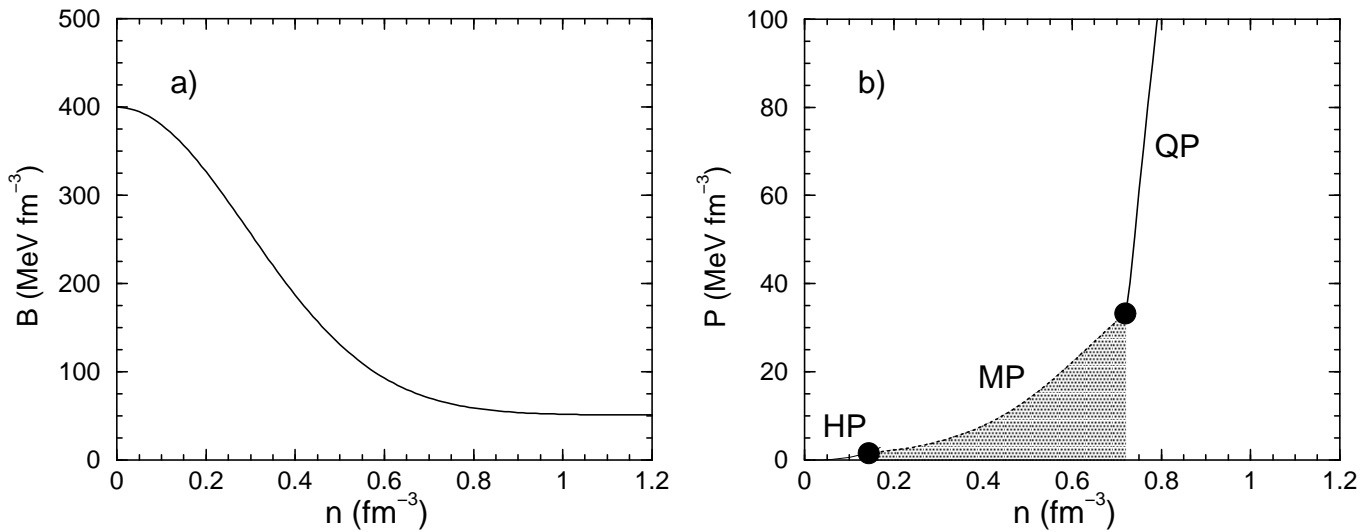}}
\caption{Bag constant $B$ versus baryon number density for a Gaussian 
  parametrization (panel a). Calculations are performed for a value of the QCD
  coupling constant of $\alpha_s = 0$.  In panel b) the total EOS including
  both hadronic and quark components is displayed. The shaded region, bordered
  by two dots, indicates the mixed phase (MP) of quarks and hadrons, while HP
  and QP label the portions of the EOS where pure hadron or pure quark phases
  are present.}
\end{figure*}

We now turn to the description of the deconfined quark phase. To describe this
phase, we have used the MIT bag model (Chodos et al.\ 1974).  In this model,
the total energy density is the sum of a non-perturbative energy shift $B$,
the bag constant, and the kinetic energy for non-interacting massive quarks of
flavors $f$ with masses $m_f$ and Fermi momenta $k_F^{(f)}$ [=$(\pi^2
n_f)^{1/3}$, where $n_f$ denotes the density of quark flavor $f$]:
\begin{equation}
{E \over{V}}  = B +   
\sum_f {3 m_f^4 \over {8\pi^2}}\Big [ x_f (2x_f^2 + 1) 
\sqrt{ x_f^2 + 1} - {\rm arsinh} \; x_f \Big ] 
\end{equation}
\noindent
with $x_f = k_F^{(f)}/m_f$. The $u$ and $d$ quarks are considered to be
massless, whereas the $s$ quark mass is chosen to be 150 MeV.  In the original
MIT bag model, the bag constant $B$ has a value of $B \approx 55\,\rm
MeV\,fm^{-3}$, which is quite small when compared with those ($\approx
210\,\rm MeV\,fm^{-3}$) estimated from lattice calculations (Satz 1982). In
this sense $B$ can be considered as a free parameter of the model.  

In an
earlier work, Burgio et al.\ (2002a) have proposed a method to constrain $B$
by experimental data obtained at the CERN SPS, where several experiments
using high energy beams of Pb nuclei reported (indirect) evidence for the
formation of a quark-gluon plasma (Heinz 2000). According to the analysis of
those experiments, the quark-hadron transition takes place at about seven
times normal nuclear matter energy density ($\epsilon_0 \approx 156\,\rm MeV\,
fm^{-3}$). By assuming that the transition to the quark-gluon plasma is
determined by the value of the energy density only, Burgio et al.\ (2002a)
found that a density dependent bag parameter is needed in order to reproduce
correctly the experimentally found value of the energy density at the
hadron-quark transition point.  More specifically, a Gaussian-like or a
Woods-Saxon-like density dependence of $B$ has been assumed, and parameters
were fitted in order to correctly reproduce the experimental
phase transition point. More details are given in (Burgio et al.\ 2002a,
2002b). We display in Fig.\ 2, panel (a), the bag constant $B$ as function of
the baryon density $n$ in the case that the assumed parametrization is a
Gaussian.

Once the density dependence of $B$ is established, the composition of
electrically charge neutral, $\beta$-stable quark matter can be determined.
The density range over which both hadronic matter and quark matter coexist is
computed as outlined in (Glendenning 1992) for a system characterized by two
conserved charges. The hadronic phase and the quark phase are allowed to be
separately electrically charged while, at the same time, total charge
neutrality is preserved. The pressure in the two phases is the same to ensure
mechanical stability, while the chemical potentials of the different species
are related to each other through baryon number conservation and electric
charge conservation.
\begin{figure*}
\centering
\resizebox{\hsize}{!}{\includegraphics[angle=270]{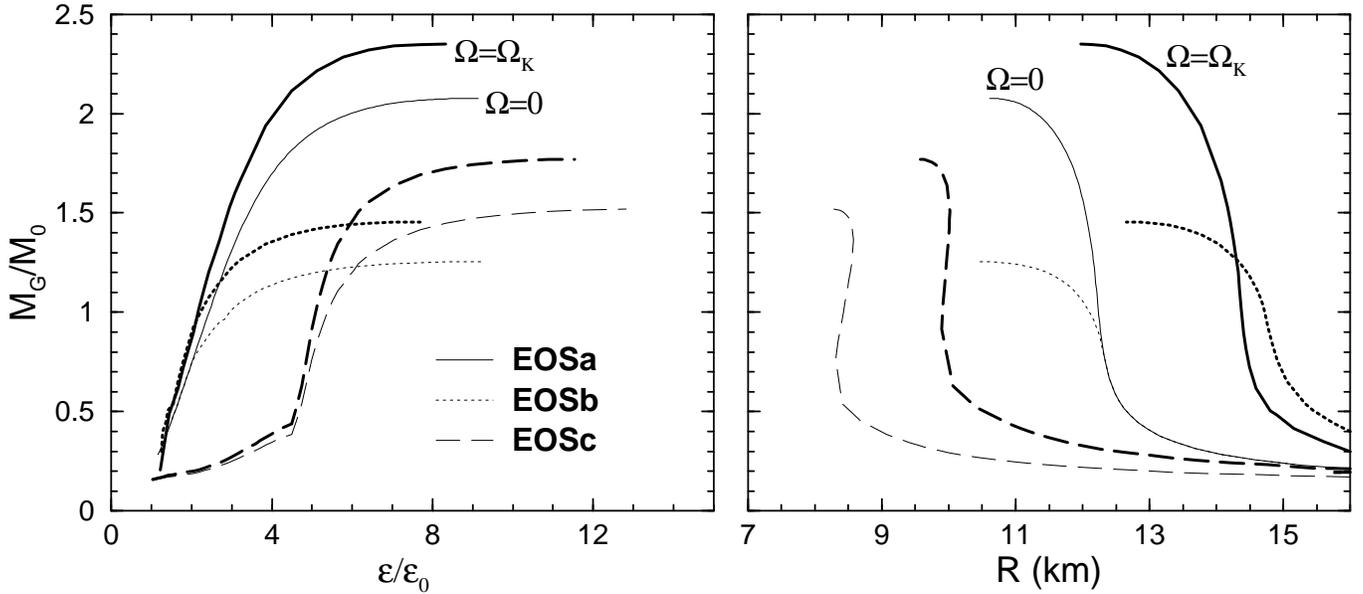}}
\caption{The gravitational mass (in units of the mass of the sun, $M_\odot$) 
  versus the normalized central energy density (left panel) and versus
  the equatorial radius (right panel). The thin lines represent static
  equilibrium configurations, whereas the thick lines display configurations
  rotating at their respective Kepler frequencies. Several different stellar
  matter compositions are considered (see text for details).}
\end{figure*}

The resulting EOS for neutron star matter, for the Gaussian-like bag
parametrization, is displayed in Fig.\ 2, panel (b), where the shaded area
indicates the mixed phase region. A pure quark phase is present at densities
above the shaded area and a pure hadronic phase is present below this area.
Moreover, the onset density of the mixed phase turns out to be slightly
smaller than the density for hyperon formation in pure hadronic matter. Of
course hyperons are still present in the hadronic component of the mixed
phase.

\section{Stellar structure equations of rotating stars}

The construction of general relativistic rotating neutron star models is a
complicated task (see, for instance, Butterworth \& Ipser 1976; Friedman,
Ipser, \& Parker 1986, 1989; Komatsu, Eriguchi, \& Hachisu 1989; Lattimer et
al.\ 1990; Salgado et al., 1994). We adopt Hartle's method (Hartle 1967;
Hartle \& Thorne 1968) to achieve this goal. The metric has the form
\begin{equation}
ds^2 = -e^{2\nu} dt^2 + e^{2\psi} (d\phi - \omega dt)^2 +
e^{2\mu} d\theta^2 + e^{2\lambda} dr^2 + O(\Omega^3) ,
\label{e:ds2}
\end{equation}
where $\nu,~ \psi,~ \mu, ~ \lambda$ denote metric functions and $\omega$ is
the angular velocity of a local inertial frame.  The general relativistic
Kepler frequency, denoted by $\Omega_K$, is given as a solution of (Friedman
et al., 1986; Weber 1999a,b)
\begin{eqnarray}
\Omega & = & e^{\nu - \psi} v(\Omega) + \omega(\Omega), \label{e:om} \\ 
v(\Omega)& = & \frac{\omega'}{2\psi'} e^{\psi - \nu} 
+ \sqrt{
\frac{\nu'}{\psi'} + \Big( \frac{\omega'}{2\psi'}e^{\psi-\nu} \Big)^2} ,
\label{e:v}
\end{eqnarray}
where $v$ denotes the orbital velocity at the star's equator. Primes refer to
derivatives with respect to the radial coordinate. All quantities in Eqs.\ 
(\ref{e:om}) and (\ref{e:v}) are to be evaluated at the star's equator.
Equations~(\ref{e:ds2}) to (\ref{e:v}) are to be solved self-consistently for a
given model for the EOS. Details can be found in Weber (1999a).

\section{Results and discussion}
In Figure 3 we plot our results obtained for non-rotating neutron stars
(thin lines) as well as for neutron stars rotating at their
self-consistent Kepler frequencies (thick lines). The vertical axes display the
gravitational mass (in units of solar masses) as a function of central star
density, $\epsilon_c$, normalized to the saturation value of infinite nuclear
matter, $\epsilon_0~ =~ 156~ \rm MeV~fm^{-3}$.  On the right-hand side in
Fig.\ 3 we illustrate the gravitational mass as function of the equatorial
radius.  The solid curves show results for neutron star matter composed of only
nucleons and leptons (EOSa). The dotted lines represent equilibrium 
configurations of neutron stars composed of nucleons, hyperons, and leptons.
The dashed curves display equilibrium configurations of hybrid stars, 
i.e., neutron stars composed of hadronic matter in the outer layers, 
followed by a mixed phase of quarks and hadrons at higher densities, 
and eventually a pure quark matter core in their centers (EOSc).
 
\begin{table}[b]
\begin{center}
\begin{tabular}{|l|c|c|c|c|}
\hline
 EOS &  $\Omega~ (10^4~{\rm s}^{-1})$ & $M_G/M_\odot$ 
     & $\epsilon_c/\epsilon_0$ & 
$R_{\rm eq}$ (km)  \\
\hline
\hline
EOSa  & 1.195 & 2.35 & 8.33  & 11.97  \\
      & 0     & 2.06 & 8.97  & 10.50  \\
\hline
EOSb  & 0.88 & 1.45 & 7.7  & 12.60  \\
      & 0    & 1.26 & 8.97 & 10.46  \\
\hline
EOSc  & 1.436 & 1.77 & 11.54 & 9.60  \\
      & 0     & 1.52 & 12.8  & 8.28 \\
\hline
\end{tabular}
\end{center}
\caption{Calculated properties of non-rotating ($\Omega=0$) and rotating
($\Omega=\Omega_K$) limiting-mass neutron star models.}
\label{table}
\end{table} 

Our results for the gravitational mass, normalized central density, equatorial
radius, and Kepler frequency for the maximum-mass star of each stellar
sequence are displayed in Table~1.  One sees an enhancement of the
gravitational mass for rotating configurations, relative to the spherical star
model of the same central energy density. This enhancement lies in the range
between $\approx 14\%$ and $17\%$, depending on the underlying EOS.
Correspondingly, for the same value of the gravitational mass, the rotating
configurations exhibit reduced central densities compared to the corresponding
non-rotating stars, which is due to the effect of the centrifugal force,
which effectively stiffens the equation of state.  Moreover, the mass increase
is accompanied by a relatively large increase of the equatorial radius, due to
the rotational deformation of the star.

A striking difference between the purely hadronic stars (EOSa and EOSb) and
the hybrid stars (EOSc) concerns their compactness. In fact, for a fixed
gravitational mass, hybrid stars are characterized by a larger central energy
density and a smaller radius than their hadronic counterparts. This holds for
both static as well as rotating configurations, as one would expect.  The
compactness of a star is crucial for its Kepler frequency as discussed next.

\begin{figure} 
\centering
\resizebox{\hsize}{!}{\includegraphics[angle=270]{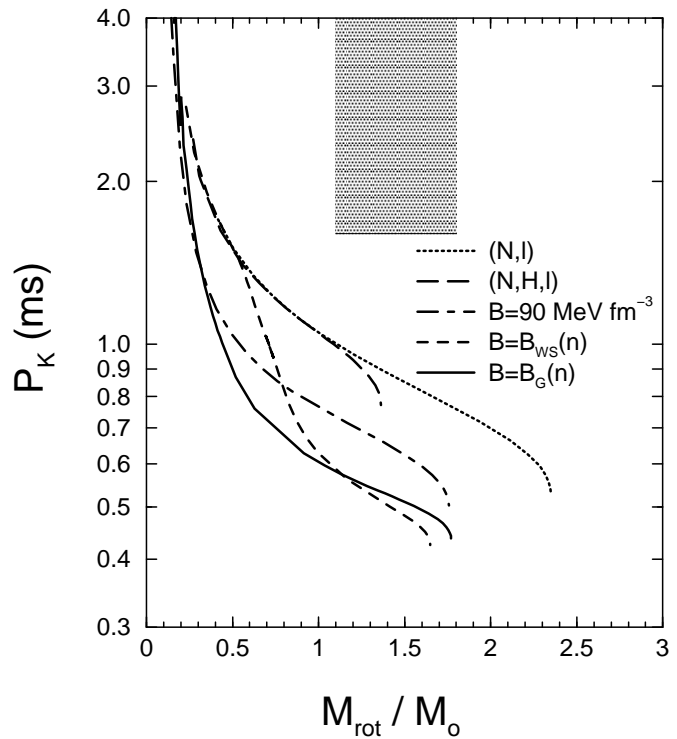}}
\caption{Kepler period versus the rotational mass for
  purely hadronic stars as well as hybrid stars.  The following core
  compositions are considered: i) nucleons and leptons (dotted line), ii)
  nucleons, hyperons, and leptons (long-dashed line), iii) hadrons and quarks
  [solid line: $B=B_G(n)$, dot-dashed line: $B=90~ \rm MeV~fm^{-3}$, 
   dashed line: $B=B_{WS}(n)$]. The shaded area represents the current
    range of observed data.}
\label{fig:P_K}
\end{figure}

In Fig.~\ref{fig:P_K} we display the Kepler periods $P_K$ ($= 2\pi /\Omega_K$)
in milliseconds versus the rotational star mass for several different stellar
sequences based on different EOSs.  The shaded area represents the current
range of the observed periods and masses.

Since the observed periods are larger than the Kepler values, which set an
absolute limit on stable rapid rotation, all EOSs studied in this paper are
compatible with the observed data. Stars made up of chemically equilibrated
nucleons or nucleons plus hyperons, shown by the dotted and long-dashed lines
respectively, show instability against mass shedding first because of their
relatively large equatorial radii.  In both cases, the limiting mass
configurations are characterized by values of the Kepler period larger than
half a millisecond, in agreement with results usually found in the literature
(Weber 1999a,b). This is different for the Kepler periods computed for the
hybrid stars in Fig.~\ref{fig:P_K} 
[labeled $B=B_G(n)$, $B=90~ \rm MeV~fm^{-3}$, and
$B=B_{WS}(n)$] which  can be even smaller than half a
millisecond. An example are the configurations along the solid line near the
mass limit, which have been obtained using a Gaussian parametrization of the
bag constant. We stress that this result does not depend on the particular
parametrization chosen for the bag constant $B$. In fact, stellar
configurations obtained with a Woods-Saxon-like bag parametrization (dashed
line) show the same behavior.

For completeness, we show in Fig.~\ref{fig:P_K} also the Kepler periods of
hybrid stars computed for a quark matter EOS with a constant bag constant of
$B = 90~\rm MeV~fm^{-3}$. In this case, no equilibrium configuration can
rotate at Kepler periods smaller than half a millisecond.  This is due to the
fact that the increase of pressure with baryon density in the pure quark phase
is less pronounced than in the case of a density dependent bag value, as shown
in (Burgio et al., 2002a).  Therefore, we deduce that only hybrid stars with a
quark phase constructed by using a density-dependent bag constant can rotate
stably at periods in the half millisecond regime. The Kepler periods of such
hybrid stars therefore rival those of absolutely stable strange stars 
(Glendenning 1990), which were thought to be the only configurations able to
withstand stable rotation down to half millisecond periods.

\section{Summary}

In this work, we have explored the Kepler periods of sequences of general
relativistic, rotating neutron and hybrid stars. For this purpose we have
employed three different equations of state derived by describing the hadronic
phase in the framework of the many-body Brueckner-Bethe-Goldstone theory using
state-of-the-art nucleon-nucleon interactions, and the quark phase within the
MIT bag model supplemented by a density dependent bag constant $B$, as
suggested by experiment.  In essence, we have found that neutron stars made of
purely hadronic matter (nucleons and hyperons) have minimum rotational periods
that are clearly larger than half a millisecond.  This may be different for
hybrid stars, i.e.,  neutron stars containing deconfined quark matter cores. In
the latter case, a density dependent bag constant provides additional
compactness to hybrid stars so that Kepler periods down to half a millisecond
are reached. Such small rotational periods are not obtained if the bag
parameter is taken to be constant, as usually assumed in the studies 
published in the literature.

We thus conclude that very rapidly rotating pulsars, which are being actively
searched for by upgraded radio telescopes and the latest generation of X-ray
satellites, may be interpreted as either strange quark matter stars or
neutron stars with deconfined quark matter interiors.

\begin{acknowledgements}
  We warmly thank Marcello Baldo (INFN Sezione di Catania) for the fruitful
  discussions about the hadron-quark phase transition in neutron stars.
\end{acknowledgements}

\end{document}